# Paradigm shift in electron-based crystallography via machine learning

Kevin Kaufmann[1], Chaoyi Zhu[2]*, Alexander S. Rosengarten[1]*, Daniel Maryanovsky[3], Tyler J. Harrington[2], Eduardo Marin[1], and Kenneth S. Vecchio[1,2]

[1]Department of NanoEngineering, UC San Diego, La Jolla, CA 92093, USA
[2]Materials Science and Engineering Program, UC San Diego, La Jolla, CA 92093, USA
[3]Department of Cognitive Science, UC San Diego, La Jolla, CA 92093, USA
*These authors contributed equally to this work.

## Abstract

Accurately determining the crystallographic structure of a material, organic or inorganic, is a critical primary step in material development and analysis. The most common practices involve analysis of diffraction patterns produced in laboratory X-ray diffractometers, transmission electron microscopes, and synchrotron X-ray sources. However, these techniques are slow, require careful sample preparation, can be difficult to access, and are prone to human error during analysis. This paper presents a newly developed methodology that represents a paradigm change in electron diffraction-based structure analysis techniques, with the potential to revolutionize multiple crystallography-related fields. A machine learning-based approach for rapid and autonomous identification of the crystal structure of metals and alloys, ceramics, and geological specimens, without any prior knowledge of the sample, is presented and demonstrated utilizing the electron backscatter diffraction technique. Electron backscatter diffraction patterns are collected from materials with well-known crystal structures, then a deep neural network model is constructed for classification to a specific Bravais lattice or point group. The applicability of this approach is evaluated on diffraction patterns from samples unknown to the computer without any human input or data filtering. This is in comparison to traditional Hough transform electron backscatter diffraction, which requires that you have already determined the phases present in your sample. The internal operations of the neural network are elucidated through visualizing the symmetry features learned by the convolutional neural network. It is determined that the model looks for the same features a crystallographer would use, even though it is not explicitly programmed to do so. This study opens the door to fully automated, high-throughput determination of crystal structures via several electron-based diffraction techniques.

# Main text

Identifying structure is a crucial step in the analysis of proteins[1–3], micro-[4,5] and macro-molecules[6], pharmaceuticals[7], geological specimens[8], synthetic materials[9–11], and many other fields. In the materials science realm, it is well known that the crystal structure plays a significant role in the properties exhibited[12,13]. Determining the complete crystal structure (i.e. crystal symmetry) of the phases in a material can be a challenging task, especially for multi-phase materials or those with low symmetry phases. Currently, the most common techniques involve either X-ray diffraction (XRD), via laboratory diffractometers or synchrotron radiation sources, or transmission electron microscopy (TEM)-based convergent beam electron diffraction (CBED)[14–16]. XRD is the faster of these two techniques, requiring only a powder or polished bulk sample, a few hours to collect diffraction intensities over a range of angles, and the time for pattern refinement to match the experimentally collected pattern to those in a database or theoretical model. However, this process is subject to misclassification by users, owing to shifts in lattice parameter, the overlapping nature of XRD peaks in multi-phase samples, texture effects, and the thresholds set by researchers for what qualifies as a 'match'. TEM studies employing convergent beam electron diffraction (CBED) are more precise than XRD in their ability to pinpoint the location of individual crystals, produce singular diffraction patterns for a given phase, and capture subtle symmetry information; however, CBED has limitations in sample preparation and the rate of data collection and analysis, and requires significant operator expertise[17–19].

A scanning electron microscope (SEM), equipped with an electron backscatter diffraction (EBSD) system, has become an essential part of characterization for crystalline materials and geological samples[8]. The discovery of electron backscatter diffraction patterns was reported by Nishikawa and Kikuchi in 1928[20]. The emergence of commercial EBSD systems can be attributed to the early research by Alam *et al.*[21], Venables *et al.*[22], Dingley[23], and the development of fully automated image analysis methods in the early 90s[24,25]. Since the introduction of automated EBSD, commercial software and hardware have evolved to capture more than 3000 patterns per second, which expands the applicability of the technique to assist researchers with more complex problems[26]. For example, the high-throughput capability of a modern EBSD system enables determination of fine-scale grain structures, sample texture, point-to-point crystal orientation, residual stress/strain, geometrically necessary dislocation densities, etc.[27–31] The relative ease of sample preparation compared with TEM samples, as well as analysis of a significantly larger sample area in less time makes SEM-EBSD an attractive technique to study location specific orientation with high precision (~2°), misorientation resolution (0.2°) and a spatial resolution ~40 nm[32]. Combined with other analytical techniques such as energy dispersive X-ray spectroscopy (EDS) or wavelength dispersive X-ray spectroscopy (WDS), phase identification is also possible[33–35], given that the chemical and structural information of the phase exists in a theoretical model or crystal database, such as the Inorganic Crystal Structure Database* (ICSD). Additionally, a method for determination of an unknown Bravais lattice using a single electron backscatter diffraction pattern, without EDS data, has also been recently proposed[36]. However, this technique requires hand-drawn lines to be overlaid with a high degree of accuracy on individual Kikuchi bands. This results in a slow and tedious methodology that requires manual annotation of each individual pattern.

However, electron backscatter diffraction (EBSD) has historically been limited to elucidating orientation of user-defined crystal structures (i.e. the crystal structure is known or assumed and only its orientation determined in EBSD). This is achieved utilizing the Hough transform, an image processing technique for finding the intersection of diffraction maxima (Kikuchi lines) in the image, and a look-up table of the orientations for only the user-defined phases (typically less than 5 phases). Furthermore, the Hough transform is unable to differentiate different crystal structures with similar symmetries or specific orientations that produce a similar Hough transform[37]. Therefore, the state-of-the-art Hough transform EBSD method requires a large amount of knowledge about the sample and/or a highly experienced user for it to be utilized in its current form. Simply put, the Hough transform approach to EBSD is incapable of

---

'determining' crystal structure and can only determine the orientation of a known crystal structure, which is preselected by the user.

Recently, the materials science field has begun to embrace the big data revolution[38]. This development has sparked an interest in machine learning, mainly for application in the discovery of new materials. Researchers have shown the ability to predict new compositions for bulk metallic glasses[39], shape memory alloys[40], Heusler compunds[41], and ultraincompressible superhard materials[42]. Other groups are developing machine learning methods to establish structure-property linkages[43,44], or predict the crystal stability of new materials[45]. There also exists a few machine learning tools for analysis of fabricated alloys. Holm *et al*.[46,47], have demonstrated the classification of 105 optical microscopy images into one of seven groups with greater than 80% accuracy[46], as well as microconstituent segmentation using the PixelNet convolutional neural network (CNN) architecture trained on manually annotated micrographs of ultrahigh carbon steel[48]. These machine learning driven analysis techniques represent important developments in the materials science toolbox, but are not broadly applicable. Optical images of microstructure are often easily confused in the real world and do not provide information about crystal structures or chemistry. The PixelNet example is highly application specific and is only applicable to high-throughput segmentation and classification of the material space it is trained on. The need for more broadly applicable analysis tools remains, and the development of a machine learning methodology for the identification of crystal symmetry from diffraction patterns demonstrates an approach broadly applicable to multiple diffraction-based analyses.

Herein, we demonstrate a novel methodology, EBSD coupled with a machine learning algorithm, to identify the origin Bravais lattice or point group of a sample from diffraction patterns. This hybrid approach to identify the underlying structure of a material in electron-based diffraction techniques functions without the need for highly skilled operators or human-based indexing approaches. The trained machine learning model is subsequently applied to materials it was not trained on, but which contain the same crystal symmetry, and identifies the correct Bravais lattice or point group with a high degree of accuracy (greater than 90% overall). This can be applied to autonomously generate crystal structure phase maps of unknown single or multi-phase materials without any prior knowledge. Moreover, the ability for machine learning to glean the symmetry features associated with crystal structure is established by monitoring the activations of the trained model. The high degree of success with determination of specific Bravais lattices, leads to the exploration of this novel methodology's ability to distinguish between structures based on their crystal point groups. Furthermore, the deep learning methodology employed herein could be applied to other disciplines outside of the materials science field including geology, pharmacology, and even structural biology[49].

**The classification model.** Having collected approximately 400,000 EBSD patterns from 40 different materials, it is necessary to classify each of these images based solely on the information contained in the image file. The classical computer vision approach is to manually engineer features and use a discriminatory model to make the ultimate decision about the Bravais lattice to which the source image belongs. Such an approach would require a multitude of heuristics—such as detecting Kikuchi bands, accounting for their slight orientation changes, looking for symmetry in the image, etc., and carrying the burden of developing the logic that defines these abstract qualities. The process of generating the corresponding computer logic is even more challenging when one considers the smallest changes in diffraction patterns resulting from orientation changes or minute atomic position differences and defects in materials of the same crystal structure. Thus, the traditional computer vision approach becomes less feasible as the number of categories grows, cannot be generalized to other crystal classes (including expanding these capabilities to point and space groups), and lacks a procedure to systematically improve prediction capabilities. On the other hand, one could use a technique that determines its own internal representation of the data so long as it performs well at the discrimination task. This is the underlying principle behind deep representation learning (i.e. deep neural networks)[50]. Such methods allow the model to find patterns that may be unintuitive or too nuanced for humans to discern. Other discovered features might be obvious to experts, but difficult to translate into specific logic.

These deep learning systems take in raw data and automatically discover, through filters learned via backpropagation[51], the abstract representations that maximize classification performance. In convolutional neural networks, such as the ones used in this work, the early layers typically learn to look for the presence or absence of edges or curves, while the later layers assemble these motifs into representative combinations and eventually familiar objects. In this case, these more familiar objects are the symmetry elements present in the diffraction pattern encoding a Bravais lattice. A schematic representation of the CNN utilized in this work is shown in Fig. 1. In the first step, a learnable filter is convolved across the image, and the scalar product between the filter and the input at every position, or 'patch', is computed to form a feature map. This is called a convolutional layer. Next, a series of alternating convolutional and pooling layers are stacked sequentially. The units in a convolutional layer are organized in feature maps, and each feature map is connected to local patches in the previous layer through a set of weights called a filter bank. All units in a feature map share the same filter banks (also called kernels), while different feature maps in a convolutional layer use different filter banks. Pooling layers are placed after convolutional layers to down sample the feature maps and produce coarse grain representations and spatial information about the features in the data. The key aspect of deep learning is that these layers of feature detection nodes are not programmed into lengthy scripts or hand-designed feature extractors, but instead 'learned' from the data. In this case, such motifs encode the underlying crystallographic symmetry present in the diffraction patterns, as demonstrated below, by investigating the learned features of the convolutional neural network.

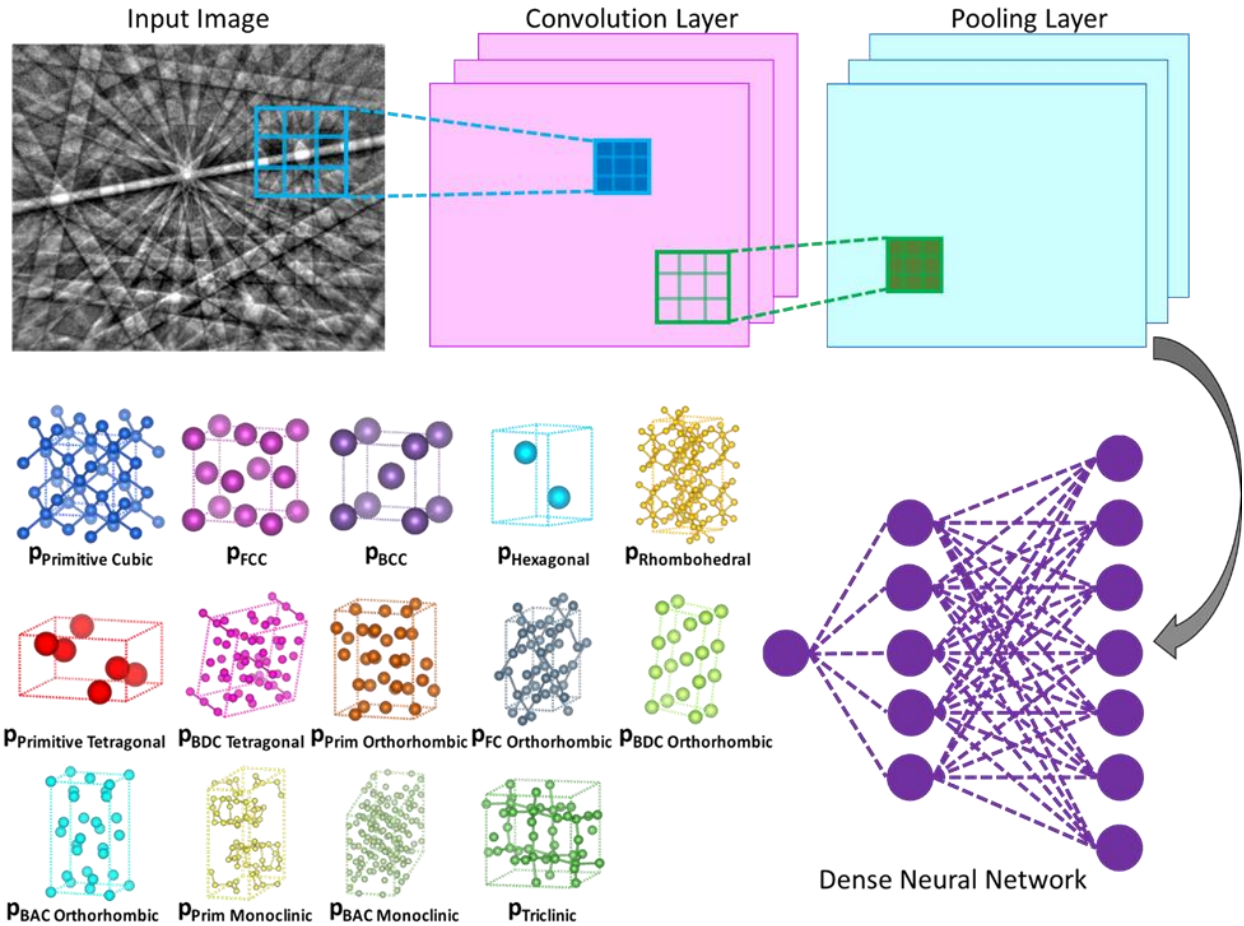

Fig. 1 | Illustration of the inner workings of a convolutional neural network. Convolutional neural networks are composed of a series of alternating convolutional and pooling layers. Each convolutional layer extracts features from its preceding layer, using filters (or kernels) learned from training the model, to form feature maps. These feature maps are then down-sampled by a pooling layer to exploit data locality. A traditional dense neural network, a simple type of classification network, is placed as the last layer of the CNN, where the probability that the input diffraction pattern belongs to a given class (e.g. Bravais lattice or point group) is computed.

**Performance of the model**. The trained model was first applied to new diffraction patterns (meaning the computer has never seen these EBSD patterns previously, and the patterns are a random mix of orientations, which may or may not be similar to the training patterns) collected from the same materials utilized for training the model via backpropagation. In Fig. 2 and Extended Data Fig. 1, both the ResNet50 and Xception architecture correctly classify nearly 300,000 diffraction patterns with a high degree of accuracy (>90% overall accuracy for each architecture). Specifically, this means the individual EBSPs were identified as belonging to one of the fourteen Bravais lattices with no user input. For further validation, EBSD patterns from nine completely different materials were selected for blind testing of the machine learning approach to autonomous identification of crystal structures. Fig. 3 and Extended Data Fig. 2 show modified confusion matrices, where correct classification is shown in the green squares rather than along the diagonal, and utilizes more than 50,000 diffraction patterns to highlight the capabilities of the proposed methodology. Each architecture correctly classifies the Bravais lattice of the unknown material with 93.5% and 91.2 % overall accuracy for ResNet50 and Xception, respectively. It is observed that the base-centered monoclinic crystal structure has a propensity to be incorrectly classified as primitive orthorhombic or rhombohedral. Further investigation as to the cause of this misclassification reveals that the base-centered monoclinic, primitive orthorhombic, and rhombohedral Bravais lattices belong to the $2/m$, $mmm$, and $\bar{3}m$ point group, respectively.

Fig. 2 | Confusion matrix displaying the model's classification results for the 14 Bravais lattices. A new set of diffraction patterns from fourteen of the materials were classified by the trained model with ResNet50 architecture. The diagonal (blue shaded boxes) in these tables represent the successful matching of the CNN predictions to the true Bravais lattices of the sample.

| True class \ Predicted | PC | FCC | BCC | Hex | Rhom | PT | BCT | PO | FCO | BCO | BaCO | PM | BaCM | Tri | | | Material |
|---|---|---|---|---|---|---|---|---|---|---|---|---|---|---|---|---|---|
| Primitive Cubic | 7368 | 4 | 2 | 6 | 22 | 10 | 0 | 3 | 45 | 2 | 424 | 14 | 2 | 12 | 93.1% | 6.9% | $Mo_3Si$ |
| Face-Centered Cubic | 2 | 8001 | 7 | 0 | 11 | 1 | 0 | 60 | 0 | 0 | 10 | 59 | 1 | | 98.1% | 1.9% | Al |
| Body-Centered Cubic | 20 | 50 | 14463 | 196 | 71 | 75 | 83 | 2 | 1 | 9 | 86 | 14 | 38 | 0 | 95.7% | 4.3% | Ta |
| Hexagonal | 0 | 3 | 2 | 5896 | 10 | 7 | 0 | 0 | 3 | 4 | 6 | 9 | 0 | 2 | 99.2% | 0.8% | Ti |
| Rhombohedral | 68 | 9 | 11 | 12 | 4548 | 189 | 4 | 5 | 75 | 5 | 28 | 52 | 13 | 53 | 89.7% | 10.3% | Ilmenite |
| Primitive Tetragonal | 2 | 5 | 8 | 3 | 6 | 8542 | 0 | 0 | 0 | 0 | 28 | 12 | 29 | | 98.9% | | Sn |
| Body-Centered Tetragonal | 0 | 2 | 2 | 1 | 16 | 0 | 12622 | 1 | 0 | 6 | 0 | 0 | 4 | 0 | 99.8% | 0.2% | Anatase |
| Primitive Orthorhombic | 0 | 0 | 0 | 3 | 3 | 1 | 0 | 3034 | 0 | 1 | 6 | 0 | 0 | 1 | 99.5% | 0.5% | Enstatite |
| Face-Centered Orthorhombic | 22 | 0 | 1 | 3 | 11 | 7 | 1 | 1 | 8706 | 0 | 226 | 21 | 1 | 3 | 96.7% | 3.3% | $Al_3Zr_2$ |
| Body-Centered Orthorhombic | 0 | 2 | 11 | 9 | 94 | 1 | 0 | 1 | 1 | 24637 | 2 | 0 | 1 | 0 | 99.5% | 0.5% | $MoPt_2$ |
| Base-Centered Orthorhombic | 0 | 0 | 0 | 1 | 0 | 19 | 0 | 53 | 0 | 0 | 6628 | 86 | 0 | 0 | 97.7% | 2.3% | $AuSn_4$ |
| Primitive Monoclinic | 1 | 0 | 0 | 0 | 1 | 1 | 2 | 1 | 0 | 0 | 0 | 5481 | 34 | 29 | 98.7% | 1.3% | Malachite |
| Base-Centered Monoclinic | 1 | 41 | 7 | 0 | 9 | 5 | 0 | 0 | 0 | 1 | 58 | 9043 | 34 | | 96.3% | 1.7% | $Fe_4Al_{13}$ |
| Triclinic | 2 | 0 | 0 | 0 | 7 | 4 | 0 | 0 | 2 | 0 | 0 | 14 | 10 | 3544 | 98.9% | 1.1% | Labradorite |

As seen in Extended Data Table 1, the 2/m and mmm point groups each only have 2-fold axis symmetry, mirror plane symmetry, and inversion center symmetry. The rhombohedral $\bar{3}$m point group shares these same symmetry elements, with the addition of one 3-fold axis symmetry. This high level of attention to specific symmetry elements present in the diffraction pattern strongly suggests that this methodology is capable of classification beyond Bravais lattices into point groups and perhaps space groups. The practicality of point group level classification is demonstrated in the following section.

**Fig. 3 | Performance of the machine learning models on data from new materials.** The ResNet50 convolutional neural network architecture perform exceedingly well on electron backscatter diffraction patterns collected from materials not used to train the model. Correct classification is identified by the green squares instead of along the diagonal.

**Classification of point groups.** The previous results allude to the possibility of point group determination in EBSD in an autonomous and high-throughput manner via machine learning. In Fig. 4 and Extended Data Fig. 3, point group classification is demonstrated for a disordered and ordered atomic arrangement of the primitive cubic, face-centered cubic, and body-centered cubic lattices. This represents a significant development in the capabilities of the EBSD technique, as well as strong evidence that the machine learning approach developed and presented herein will be capable of at least point group level classification in other structure probing techniques.

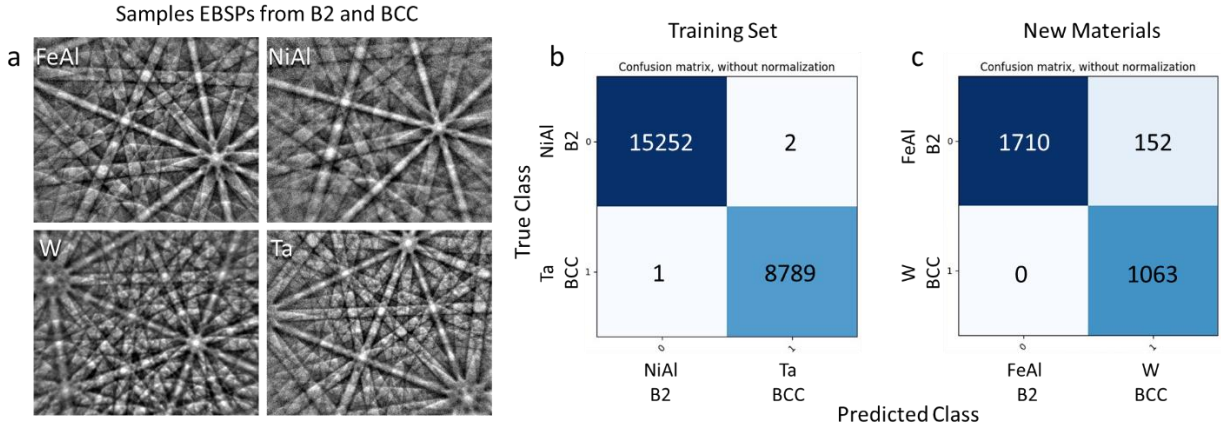

Fig. 4 | Demonstration of machine learning aided EBSD's capability to autonomously identify point groups. **a,** An electron backscatter diffraction pattern for each of the four materials. Zone axes with the same symmetry are seen in each diffraction pattern. **b,** Results of training the model to discriminate between the B2 (atomically ordered) and BCC (atomically disordered) point groups. **c,** Resultant confusion matrix after testing the model blindly on two materials from the same point groups.

**Visualizing the learned features**. The excellent performance of these deep neural networks in correctly identifying the structure of materials previously unknown to the system inherently requires an intuitive and understandable interpretation of how the model is arriving at the correct classifications. Fig. 5 elucidates an example of how the model accomplishes this by finding features in the diffraction patterns that represent the symmetry present in the origin lattice. Two diffraction patterns, one for nickel and one for aluminum, of similar crystallographic orientation, were selected for analysis of the 'importance' of the features present. The weight of local regions in the image is discovered using the neural network architecture and employing a set of tools called Grad-CAM[52]. After computing the 'importance' of these local regions in the diffraction image, the weights are normalized from 0 (dark blue) to 1 (dark red) and overlaid on the original image as a heatmap. In the example provided for diffraction patterns collected from similarly oriented crystals of nickel and aluminum, it is observed that the activation maps are remarkably similar and elucidate the intense interest of the network in symmetry located at the zone axes. For each diffraction pattern, the $[1\bar{1}2]$ and $[112]$ zone axes (2-fold symmetry) are the regions of greatest interest. The machine learning algorithm couples this information with the presence of the $[001]$ (4-fold symmetry) and $[0\bar{1}3]$ (2-fold symmetry), as well as their spatial relationship owing to pooling layers, to correctly identify the origin Bravais lattice as face-centered cubic. Extended Data Fig. 4 displays one similar heatmap for each of the 28 materials utilized in the training set. For each material, a similar interest in the information nearest the zone axes is observed.

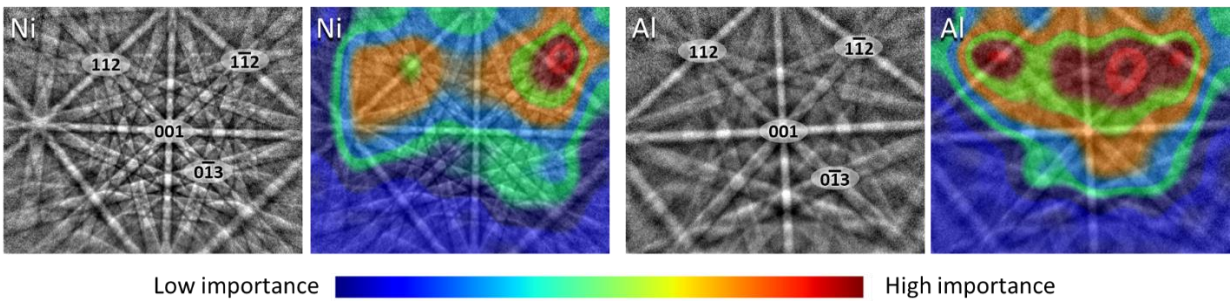

Low importance ⬛ High importance

Fig. 5 | Visualizing the features utilized for classification of a diffraction pattern to face-centered cubic (FCC). An electron backscatter diffraction pattern from nickel (Ni) and aluminum (Al) were selected from nearly identical orientations. In the diffraction patterns, four of the zone axes present in each are labeled. The corresponding heatmap displays the importance of information in the image for correctly classifying it as FCC. It is observed that for each of these two images, the symmetry information near the $[1\bar{1}2]$ zone axis produces the highest activation, followed by the $[112]$ zone axis, and symmetry shared by the $[001]$ and $[0\bar{1}3]$ zone axes.

**Phase mapping with machine learning.** To date, it has been generally accepted that computers are incapable of band detection utilizing the original diffraction patterns, hence the use of a Hough transform to locate diffraction maxima in the EBSP. However, Hough transform EBSD is limited to orientation determination of user-defined phases (i.e. the crystal structure is known or assumed and only its orientation determined). Rather than utilizing the Kikuchi diffraction lines directly, the Hough transform is utilized to locate diffraction maxima and compare a small amount of information from the transformed image to a dictionary of orientations for the small subset (typically less than 5) user-selected phases. The small subset of information being utilized often results in mis-classification of similar crystal structures and orientations that produce similar diffraction maxima in different crystal systems[37]. Therefore, the applicability of the state-of-the-art technique is limited to situations, where a great deal of information is already known about the sample, and an experienced operator is available to confirm the diffraction patterns are not being incorrectly indexed.

The methodology demonstrated herein reduces the requirements for prior knowledge of the sample and highly trained operators. Instead of relying on a subset of user selected phases, this approach autonomously utilizes all the information in each collected diffraction pattern to determine the crystal structure at that location. This advancement is demonstrated in Fig. 6 on a sample of rutilated quartz. Fig. 6b is a forward scattered electron image, which clearly demonstrates the two-phase nature of the sample, with quartz appearing recessed and rutile appearing raised above the surface. In Fig. 6a, a phase map of quartz and rutile was generated after a user selected the quartz and rutile phases to serve as the look-up table and subsequently collected diffraction patterns for Hough transform indexing. Fig. 6c was generated by collecting the diffraction patterns and allowing the convolutional neural network to determine the crystal structure, out of all the possible 14 Bravais lattices, that produced each diffraction pattern. The machine learning generated phase map is nearly identical to that generated by the state-of-the-art method, except it was generated without any human supplied information. Future combination of this technique with chemistry information and a crystal database has the potential to enable highly accurate autonomous phase identification.

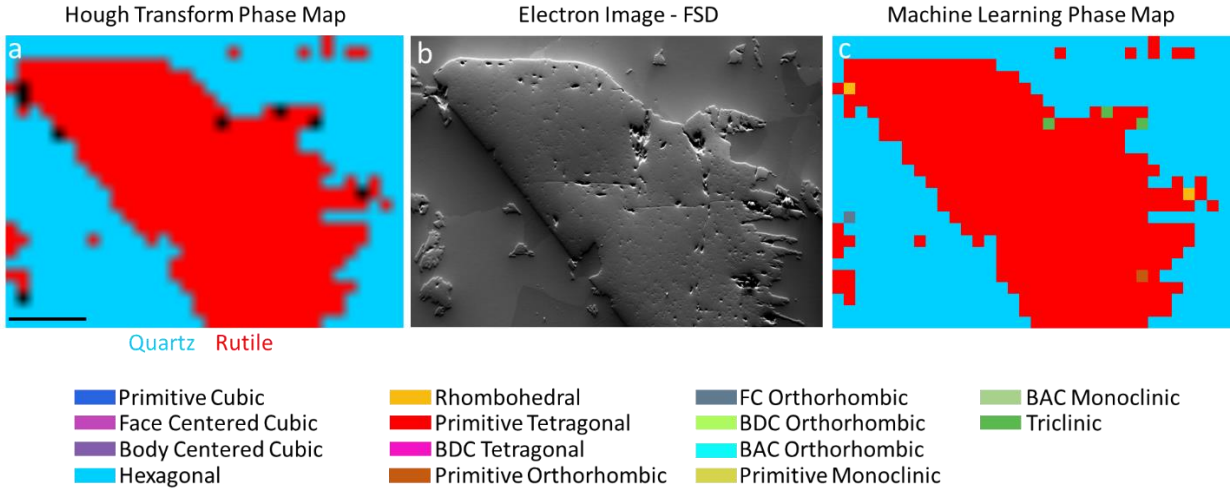

| Primitive Cubic | Rhombohedral | FC Orthorhombic | BAC Monoclinic |
| Face Centered Cubic | Primitive Tetragonal | BDC Orthorhombic | Triclinic |
| Body Centered Cubic | BDC Tetragonal | BAC Orthorhombic | |
| Hexagonal | Primitive Orthorhombic | Primitive Monoclinic | |

Fig. 6 | Utilizing machine learning to predict the number of phases and their location in the material. The phase map on the left was generated by traditional means where the user had to select quartz and rutile as the two specific phases present in the sample. The center electron image is the region of the sample from which the diffraction patterns were collected. The image on right is a phase map generated via machine learning determination of the Bravais lattice for each diffraction pattern. Black scale bar in (a) is 100μm.

## Conclusions

A revolutionary approach to identifying the underlying structure of a material in electron-based diffraction techniques has been presented here and its success demonstrated in determining the origin Bravais lattice or point group of an electron backscatter diffraction pattern. This methodology immediately enables high-throughput autonomous determination of crystal symmetry in electron backscatter diffraction and will be of great importance to multiple techniques and disciplines, including those outside of EBSD and materials science. The method is demonstrated and validated via a comprehensive set of studies to be capable of classification of diffraction patterns to the 14 Bravais lattices and the 32 point groups; areas where the state-of-the-art Hough transform based EBSD fails and only convergent beam electron diffraction, precision XRD, and synchrotron are viable techniques. Feature visualization demonstrates that the CNN identifies specific features resulting from unique crystal symmetry operations within the diffraction pattern images. This demonstrates feasibility of moving toward determination of precise space groups of unknown systems, where the only differences may be a single different symmetry operator. Moreover, we show the application of this method to phase map the Bravais lattice of a two-phase specimen without any human supplied information and compare the results with the traditional Hough transform-based approach supplied with user selected libraries. Future improvements to our methodology could be realized with the design of neural network architectures developed specifically for this task, an increased range of materials for the network to learn from, or data filters to prevent the network learning from or attempting to classify diffraction patterns without enough information present for robust classification. In the near future, we foresee this hybrid approach being utilized to develop similar revolutionary electron-based structure determination methodologies in a multitude of other crystallography-related fields including pharmacology, structural biology, and geology, and should be extendable to X-ray diffraction techniques as well.

## Methods

**Electron backscatter diffraction pattern collection**. Elements or alloys of known crystal structure were collected and polished for EBSD. High purity samples of each metal, alloy, or ceramic material ($Cr_3Si$, $Mo_3Si$, $V_8C_7$, Si, Ni, Al, NbC, TaC, W, Ta, β-Ti, Fe, WC, Ti, $Al_2O_3$, Sn, $Mo_5Si_3$, $Mo_2C$, $TiSi_2$, $Al_3Zr_2$, $Ni_2V$, $MoPt_2$, $AuSn_4$, $ZrSi_2$, $Nb_2O_5$, $Fe_4Al_{13}$, $FeNi_3$, TiC, NiAl, and $Ni_3Al$), and carefully selected geological specimens (quartz, ilmenite, rutile, anatase, forsterite, enstatite, malachite, jadeite, diopside, and anorthite) were selected. Extended Data Fig. 5 displays the materials and their Bravais lattice. In multi-phase geological specimens, only patterns identifiable to one of the phases were analyzed. EBSPs were collected in a Thermo-Fisher (formerly FEI) APREO scanning electron microscope (SEM) equipped with an Oxford Symmetry EBSD detector. Extended Data Fig. 6 details the experimental setup within the SEM chamber. The Oxford Symmetry EBSD detector was utilized in high resolution (1244x1024) mode with frame averaging. EBSPs were collected from multiple large areas with large step sizes chosen to maximize the number of differently oriented patterns taken from unique grains. After collecting high resolution EBSPs from each material, all patterns collected were exported as tiff images. Extended Data Fig. 7 displays example images of the high-resolution diffraction patterns collected. None of the collected patterns were excluded from training, testing, or validation studies unless their origin was uncertain (i.e. non-indexable patterns from multi-phase geological specimens). That is to say, the collected data was not filtered for pattern quality via any means, and the library of images for each phase may contain partial or low-quality diffraction patterns, which will decrease the accuracy of their identification.

**Neural network architecture and training procedure**. Two well-studied CNN architectures designed for broad image classification tasks, ResNet50[53] and Xception[54], were utilized in this work. Training was performed using Adam optimization[55], with batches of 32 images, and a minimum delta of 0.001 in the validation loss for early stopping criteria. The CNNs were implemented with TensorFlow[56] and Keras[57]. Refer to Extended Data Fig. 8 for a detailed schematic of the test-train workflow, and refer to Extended Data Fig. 9 for the connection between the 7 crystal structures, 14 Bravais lattices, 32 crystal point groups, and 230 crystal space groups.

For the 14 Bravais lattices, two materials from each class were selected as training materials (e.g. nickel and aluminum for face centered cubic). Two materials were selected for training to increase the network's focus on common symmetry elements in the diffraction patterns. For training the model, 640 diffraction patterns from each material were randomly selected, and the 300,000 remaining diffraction patterns from those same materials were employed as the test set to gauge the model's accuracy after training.

In the point group classification example, the ResNet50 convolutional neural network was trained as a binary classifier between an atomically ordered and disordered version of the same lattice. The three models were trained using diffraction patterns from $FeNi_3$ ($L1_2$) and $Cr_3Si$ (primitive cubic), TiC (B1) and Ni (FCC), and NiAl (B2) and Ta (BCC). The trained model was then blind tested on two materials that were new to the machine learning model: $Ni_3Al$ ($L1_2$) and $Mo_3Si$ (primitive cubic), TaC (B1) and Al (FCC), and FeAl (B2) and W (BCC).

**Validation studies.** The accuracy of the trained models was evaluated using new diffraction patterns collected from each of the materials utilized to train the deep neural network as well as diffraction patterns from materials the model had not previously encountered. Heatmap overlays giving class-specific gradient information at the classification layer of the Xception architecture were produced using Grad-CAM[52]. This technique reveals the importance of local regions in the diffraction pattern to assist in the determination of why the algorithm classified the image to a particular structure.

**Data availability.** The trained models generated during and/or analyzed during the current study are available from the corresponding author on reasonable request. The python code for implementing these models with Keras and Tensorflow is available from the corresponding author upon request. The diffraction patterns analyzed during the current study are not made available due to the sheer size of the library (nearly 3 terabytes).

**Acknowledgements**. K. Kaufmann was supported by the Department of Defense (DoD) through the National Defense Science and Engineering Graduate Fellowship (NDSEG) Program. K. Kaufmann would also like to acknowledge the support of the ARCS Foundation. C. Zhu is partially sponsored by Joint DoD/DOE Munitions Technology Development Program and the Dynamic Materials Science Campaign at LANL. The authors would like to thank Dr. Emily Chin of Scripps Institute of Oceanography and her students for their assistance collecting and identifying mineral and geological samples. KV would like to acknowledge the financial generosity of the Oerlikon Group in support of his research group.

**Author contributions.** K. K. assisted in developing the idea, performed the bulk of the experimental work, worked on later versions of the neural network python code, and prepared the initial draft of the manuscript and figures. C. Z. assisted with the development of the idea into a scientific study, contributed his knowledge of EBSD, assisted with the figures, and developed MATLAB code for phase mapping the machine learning predictions and pattern database management. A. S. R. developed the initial python code for implementing the neural networks. A. S. R. and D. M. managed the deployment of Grad-CAM tools and assisted in the analysis of the results from the deep learning models. T. J. H. assisted K. K. with materials selection, fabrication, and processing. T. J. H. also helped focus the research direction. E. M. assisted K. K. with materials fabrication, processing, and analysis. K. V. led the development of the idea, guiding the focus of the project, and reviewed and revised the manuscript. All the authors participated in analyzing and interpreting the final data. All authors contributed to the discussions and revisions of the manuscript.

**Competing interests.** The authors declare no competing interests.

**Correspondence and requests for materials** should be addressed to K. V.

**Extended Data:**

| Point Group | Lattice | 2-fold | 3-fold | 4-fold | 6-fold | Planes | Center |
|---|---|---|---|---|---|---|---|
| 1 | Triclinic | | | | | | |
| $\bar{1}$ | Triclinic | | | | | | yes |
| 2 | Monoclinic | 1 | | | | | |
| $m$ | Monoclinic | | | | | 1 | |
| $2/m$ | Monoclinic | 1 | | | | 1 | yes |
| 222 | Orthorhombic | 3 | | | | | |
| $mm2$ | Orthorhombic | 1 | | | | 2 | |
| $mmm$ | Orthorhombic | 3 | | | | 3 | yes |
| 4 | Tertragonal | | | 1 | | | |
| $\bar{4}$ | Tertragonal | 1 | | | | | |
| $4/m$ | Tertragonal | 1 | | 1 | | 1 | yes |
| 422 | Tertragonal | 4 | | 1 | | | |
| $4mm$ | Tertragonal | | | | | 4 | |
| $\bar{4}2m$ | Tertragonal | 3 | | | | 2 | |
| $4/mmm$ | Tertragonal | 4 | | 1 | | 5 | yes |
| 23 | Cubic | 3 | 4 | | | | |
| $m3$ | Cubic | 3 | 4 | | | 3 | yes |
| 432 | Cubic | 6 | 4 | 3 | | | |
| $\bar{4}3m$ | Cubic | 3 | 4 | | | 6 | |
| $m3m$ | Cubic | 6 | 4 | 3 | | 9 | yes |
| 3 | Trigonal | | 1 | | | | |
| $\bar{3}$ | Trigonal | | 1 | | | | yes |
| 32 | Trigonal | 3 | 1 | | | | |
| $3m$ | Trigonal | | 1 | | | 3 | |
| $\bar{3}m$ | Trigonal | 3 | 1 | | | 3 | yes |
| 6 | Hexagonal | | | | 1 | | |
| $\bar{6}$ | Hexagonal | | 1 | | | 1 | |
| $6/m$ | Hexagonal | | | | 1 | 1 | yes |
| 622 | Hexagonal | 6 | | | 1 | | |
| $6mm$ | Hexagonal | | | | 1 | 6 | |
| $\bar{6}m2$ | Hexagonal | 3 | 1 | | | 4 | |
| $6/mmm$ | Hexagonal | 6 | | | 1 | 7 | yes |

Extended Data Table 1 | The point groups and their respective lattices are listed with the symmetry elements that are present for the crystal structure.

**a**

| True class | Primitive Cubic | Face-Centered Cubic | Body-Centered Cubic | Hexagonal | Rhombohedral | Primitive Tetragonal | Body-Centered Tetragonal | Primitive Orthorhombic | Face-Centered Orthorhombic | Body-Centered Orthorhombic | Base-Centered Orthorhombic | Primitive Monoclinic | Base-Centered Monoclinic | Triclinic | | | |
|---|---|---|---|---|---|---|---|---|---|---|---|---|---|---|---|---|---|
| Primitive Cubic | 32294 | 9 | 2 | 0 | 26 | 36 | 1 | 1 | 32 | 0 | 0 | 11 | 3 | 57 | 99.5% | 0.5% | $Cr_3Si$ |
| Face-Centered Cubic | 25 | 11515 | 1781 | 11 | 14 | 1 | 146 | 0 | 0 | 11 | 178 | 2 | 13 | 0 | 84.1% | 15.9% | Ni |
| Body-Centered Cubic | 5 | 224 | 18983 | 451 | 43 | 0 | 26 | 9 | 8 | 1 | 2 | 0 | 5 | 0 | 96.1% | 3.9% | W |
| Hexagonal | 4 | 249 | 157 | 13134 | 6 | 23 | 17 | 3 | 8 | 133 | 7 | 11 | 5 | 47 | 95.1% | 4.9% | WC |
| Rhombohedral | 1 | 2 | 7 | 19 | 8206 | 21 | 121 | 22 | 33 | 16 | 5 | 6 | 58 | 25 | 96.1% | 3.9% | $Al_2O_3$ |
| Primitive Tetragonal | 0 | 0 | 0 | 0 | 0 | 212 | 0 | 1 | 0 | 0 | 0 | 0 | 0 | 0 | 99.1% | 0.9% | Rutile |
| Body-Centered Tetragonal | 0 | 0 | 9 | 17 | 4 | 3 | 6060 | 5 | 5 | 1 | 3 | 1 | 0 | 1 | 99.2% | 0.8% | $Mo_5Si_3$ |
| Primitive Orthorhombic | 0 | 0 | 2 | 7 | 6 | 1 | 5 | 6980 | 3 | 3 | 52 | 1 | 3 | 2 | 98.8% | 1.2% | Forsterite |
| Face-Centered Orthorhombic | 16 | 7 | 3 | 17 | 216 | 20 | 0 | 9 | 10246 | 10 | 217 | 32 | 8 | 4 | 94.8% | 5.2% | $TiSi_2$ |
| Body-Centered Orthorhombic | 0 | 7 | 0 | 6 | 23 | 6 | 0 | 37 | 2 | 7842 | 27 | 25 | 6 | 0 | 98.3% | 1.7% | $Ni_2V$ |
| Base-Centered Orthorhombic | 72 | 9 | 24 | 7 | 17 | 24 | 0 | 117 | 1 | 0 | 10442 | 32 | 1 | 1 | 97.2% | 2.8% | $ZrSi_2$ |
| Primitive Monoclinic | 0 | 0 | 0 | 7 | 7 | 12 | 0 | 3 | 5 | 0 | 33 | 7026 | 1 | 1 | 99.0% | 1.0% | $Nb_2O_5$ |
| Base-Centered Monoclinic | 23 | 11 | 3 | 4 | 1261 | 44 | 102 | 1305 | 131 | 1 | 9 | 19 | 19686 | 1614 | 81.3% | 18.7% | Jadeite |
| Triclinic | 5 | 0 | 3 | 0 | 87 | 15 | 0 | 2 | 1 | 0 | 0 | 12 | 25 | 5872 | 97.5% | 2.5% | Anorthite |

Predicted class

b

Confusion matrix — True class (rows) vs Predicted class (columns)

| True class | Primitive Cubic | Face-Centered Cubic | Body-Centered Cubic | Hexagonal | Rhombohedral | Primitive Tetragonal | Body-Centered Tetragonal | Primitive Orthorhombic | Face-Centered Orthorhombic | Body-Centered Orthorhombic | Base-Centered Orthorhombic | Primitive Monoclinic | Base-Centered Monoclinic | Triclinic | % | % | |
|---|---|---|---|---|---|---|---|---|---|---|---|---|---|---|---|---|---|
| Primitive Cubic | 7232 | 3 | 11 | 0 | 14 | 28 | 0 | 2 | 84 | 0 | 515 | 11 | 2 | 12 | 91.4% | 8.6% | Mo$_3$Si |
| Face-Centered Cubic | 13 | 7911 | 8 | 0 | 83 | 0 | 1 | 0 | 22 | 0 | 1 | 113 | 0 | | 97.0% | 3.0% | Al |
| Body-Centered Cubic | 0 | 13 | 14812 | 13 | 51 | 2 | 80 | 1 | 11 | 10 | 78 | 0 | 36 | 1 | 98.0% | 2.0% | Ta |
| Hexagonal | 0 | 3 | 7 | 8908 | 7 | 3 | 0 | 1 | 0 | 3 | 2 | 5 | 1 | 2 | 99.4% | 0.6% | Ti |
| Rhombohedral | 1 | 6 | 1 | 0 | 4522 | 361 | 0 | 4 | 73 | 2 | 9 | 16 | 8 | 57 | 89.2% | 10.8% | Ilmenite |
| Primitive Tetragonal | 0 | 3 | 6 | 0 | 3 | 8571 | 0 | 0 | 0 | 1 | 22 | 7 | 22 | | 99.3% | 0.7% | Sn |
| Body-Centered Tetragonal | 0 | 0 | 6 | 0 | 8 | 0 | 12632 | 0 | 2 | 0 | 0 | 0 | 0 | | 99.9% | 0.1% | Anatase |
| Primitive Orthorhombic | 0 | 1 | 0 | 0 | 2 | 4 | 0 | 3034 | 0 | 2 | 4 | 0 | 1 | | 99.5% | 0.5% | Enstatite |
| Face-Centered Orthorhombic | 21 | 4 | 2 | 3 | 10 | 8 | 2 | 2 | 8756 | 0 | 161 | 32 | 1 | 1 | 97.3% | 2.7% | Al$_3$Zr$_2$ |
| Body-Centered Orthorhombic | 0 | 5 | 75 | 3 | 270 | 3 | 1 | 3 | 4 | 24394 | 0 | 0 | 0 | 1 | 98.5% | 1.5% | MoPt$_2$ |
| Base-Centered Orthorhombic | 0 | 0 | 0 | 0 | 2 | 5 | 0 | 44 | 0 | 0 | 6697 | 37 | 0 | 2 | 98.7% | 1.3% | AuSn$_4$ |
| Primitive Monoclinic | 0 | 0 | 2 | 0 | 0 | 0 | 0 | 2 | 3 | 0 | 1 | 5605 | 14 | 24 | 99.2% | 0.8% | Malachite |
| Base-Centered Monoclinic | 2 | 37 | 1 | 0 | 3 | 5 | 1 | 1 | 0 | 0 | 14 | 9125 | 11 | | 99.2% | 0.8% | Fe$_4$Al$_{13}$ |
| Triclinic | 0 | 0 | 0 | 0 | 1 | 1 | 0 | 6 | 0 | 0 | 10 | 0 | 3565 | | 99.5% | 0.5% | Labradorite |

Predicted class

**c**

Confusion matrix (Xception architecture, **c**). True class (rows) vs Predicted class (columns).

| True class | Primitive Cubic | Face-Centered Cubic | Body-Centered Cubic | Hexagonal | Rhombohedral | Primitive Tetragonal | Body-Centered Tetragonal | Primitive Orthorhombic | Face-Centered Orthorhombic | Body-Centered Orthorhombic | Base-Centered Orthorhombic | Primitive Monoclinic | Base-Centered Monoclinic | Triclinic | % correct | % error | Material |
|---|---|---|---|---|---|---|---|---|---|---|---|---|---|---|---|---|---|
| Primitive Cubic | 2308 | 16 | 3 | 2 | 24 | 23 | 0 | 1 | 6 | 0 | 2 | 2 | 23 | 65 | 93.2% | 6.8% | $Cr_3Si$ |
| Face-Centered Cubic | 19 | 10853 | 587 | 90 | 1227 | 2 | 159 | 0 | 9 | 1 | 726 | 2 | 22 | 0 | 78.2% | 20.8% | Ni |
| Body-Centered Cubic | 9 | 596 | 18794 | 293 | 19 | 0 | 35 | 1 | 0 | 10 | 0 | 1 | 0 | 0 | 95.1% | 4.9% | W |
| Hexagonal | 7 | 259 | 192 | 13187 | 15 | 27 | 15 | 2 | 11 | 47 | 5 | 8 | 4 | 25 | 95.5% | 4.5% | WC |
| Rhombohedral | 3 | 8 | 5 | 3 | 8329 | 17 | 17 | 0 | 7 | 7 | 6 | 5 | 47 | 35 | 97.5% | 2.5% | $Al_2O_3$ |
| Primitive Tetragonal | 0 | 0 | 0 | 0 | 1 | 213 | 0 | 0 | 0 | 0 | 0 | 0 | 0 | 0 | 99.5% | 0.5% | Rutile |
| Body-Centered Tetragonal | 7 | 0 | 10 | 0 | 19 | 15 | 6045 | 3 | 6 | 0 | 3 | 1 | 0 | 0 | 99.0% | 1.0% | $Mo_5Si_3$ |
| Primitive Orthorhombic | 0 | 0 | 0 | 1 | 5 | 4 | 7 | 7015 | 0 | 3 | 24 | 2 | 1 | 3 | 99.3% | 0.7% | Forsterite |
| Face-Centered Orthorhombic | 9 | 21 | 3 | 17 | 166 | 0 | 6 | 0 | 10397 | 20 | 114 | 44 | 4 | 4 | 96.2% | 3.8% | $TiSi_2$ |
| Body-Centered Orthorhombic | 0 | 3 | 4 | 1 | 33 | 11 | 0 | 23 | 0 | 7872 | 20 | 9 | 3 | 2 | 98.6% | 1.4% | $Ni_2V$ |
| Base-Centered Orthorhombic | 26 | 8 | 23 | 0 | 9 | 17 | 0 | 0 | 65 | 1 | 10588 | 9 | 0 | 2 | 98.5% | 1.5% | $ZrSi_2$ |
| Primitive Monoclinic | 0 | 0 | 0 | 1 | 2 | 7 | 0 | 1 | 6 | 2 | 14 | 7054 | 3 | 5 | 99.4% | 0.6% | $Nb_2O_5$ |
| Base-Centered Monoclinic | 48 | 23 | 13 | 2 | 5968 | 9 | 1025 | 1903 | 464 | 1 | 110 | 22 | 12931 | 1704 | 53.4% | 46.6% | Jadeite |
| Triclinic | 1 | 0 | 3 | 0 | 74 | 0 | 9 | 0 | 0 | 0 | 0 | 2 | 40 | 5892 | 97.8% | 2.2% | Anorthite |

Predicted class

Extended Data Fig. 1 | Confusion matrices displaying the classification results for the 14 Bravais lattices. A new set of diffraction patterns were classified by the ResNet50 (**a**) and Xception (**b, c**) architecture. The two trained models were tested using newly collected diffraction patterns from the twenty-eight materials used to train the machine learning models. The diagonal (blue shaded boxes) in these tables represent the successful matching of the CNN predictions to the true Bravais lattices of the sample.

| True class | Primitive Cubic | Face-Centered Cubic | Body-Centered Cubic | Hexagonal | Rhombohedral | Primitive Tetragonal | Body-Centered Tetragonal | Primitive Orthorhombic | Face-Centered Orthorhombic | Body-Centered Orthorhombic | Base-Centered Orthorhombic | Primitive Monoclinic | Base-Centered Monoclinic | Triclinic | |
|---|---|---|---|---|---|---|---|---|---|---|---|---|---|---|---|
| Primitive Cubic | 7881 | 106 | 49 | 5 | 27 | 34 | 43 | 18 | 5 | 6 | 33 | 7 | 33 | 4 | 95.5% $V_8C_7$ |
| Primitive Cubic | 4906 | 1 | 4 | 0 | 2 | 0 | 0 | 2 | 133 | 7 | 1 | 0 | 4 | 0 | 97.0% Si |
| Face-Centered Cubic | 0 | 1663 | 4 | 0 | 0 | 0 | 0 | 0 | 0 | 0 | 0 | 0 | 0 | 0 | 99.8% NbC |
| Face-Centered Cubic | 1 | 2414 | 20 | 0 | 4 | 1 | 2 | 0 | 0 | 0 | 1 | 4 | 0 | 2 | 98.6% TaC |
| Body-Centered Cubic | 1 | 240 | 10956 | 0 | 27 | 230 | 0 | 0 | 0 | 0 | 5 | 2 | 46 | 2 | 95.2% Beta Ti |
| Body-Centered Cubic | 1 | 30 | 6158 | 0 | 0 | 0 | 0 | 0 | 0 | 5 | 26 | 206 | 0 | 0 | 95.8% Fe |
| Hexagonal | 0 | 0 | 0 | 2988 | 2 | 20 | 0 | 8 | 0 | 0 | 0 | 0 | 0 | 3 | 98.9% Quartz |
| Primitive Monoclinic | 1306 | 2 | 167 | 35 | 26 | 2 | 8 | 0 | 580 | 50 | 284 | 8783 | 9 | 0 | 78.1% $Mo_2C$ |
| Base-Centered Monoclinic | 18 | 0 | 0 | 0 | 6 | 2 | 2 | 613 | 12 | 14 | 22 | 1 | 0 | 0 | 0% Diopside |

Predicted class

Extended Data Fig. 2 | Performance of the machine learning model on data from new materials. The Xception convolutional neural network architecture performs exceedingly well on electron backscatter diffraction patterns collected from materials not used to train the model. Correct classification is identified by the green squares instead of along the diagonal.

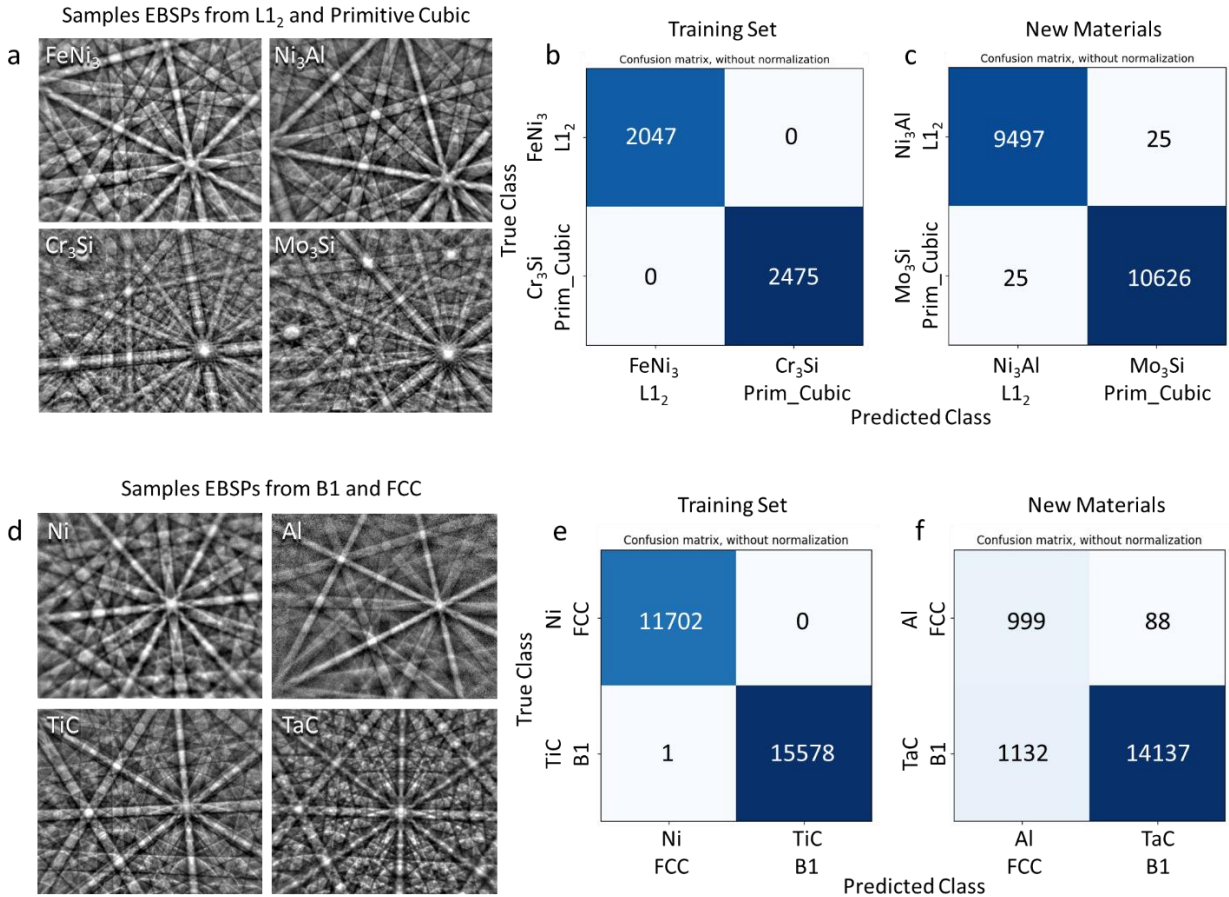

Extended Data Fig. 3 | Demonstrations of machine learning aided EBSD's capability to autonomously identify point groups in new materials. **a, d,** An example electron backscatter diffraction pattern for each of the materials. Zone axes with the same symmetry can be seen in each diffraction pattern. **b, e,** Results of training the model to discriminate between an atomically ordered and atomically disordered point group for the same crystal structure. **c, f,** Resultant confusion matrix after testing the model blindly on two materials from the same point groups.

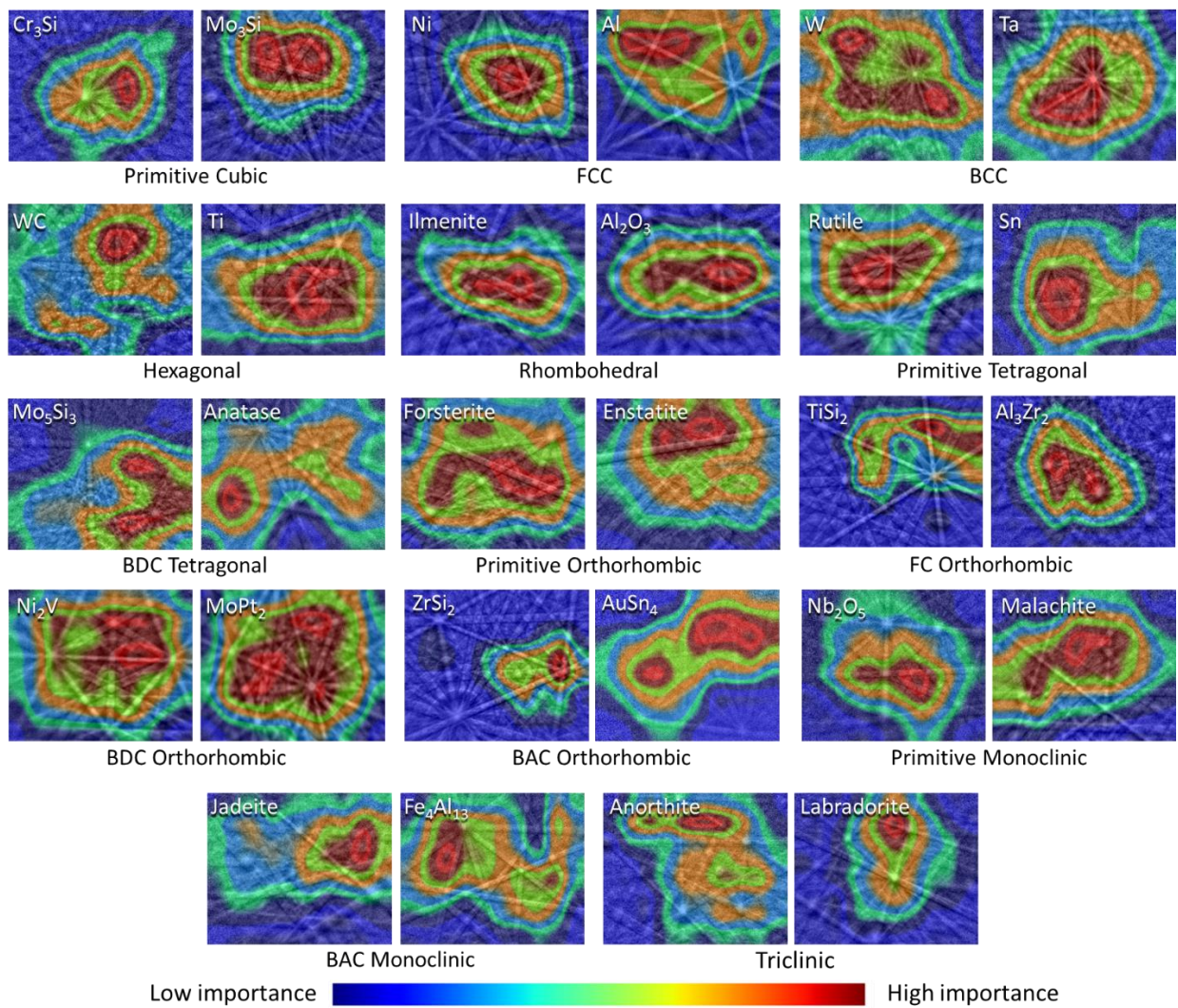

Low importance ▬▬▬▬▬▬▬ High importance

Extended Data Fig. 4 | Heatmaps elucidating the normalized importance of features (dark blue to dark red) in each diffraction pattern for classification of the correct Bravais lattice. One example is presented for each material utilized in the training set. It is observed that the symmetry features present in the image produce the highest activations for determining their origin Bravais lattice.

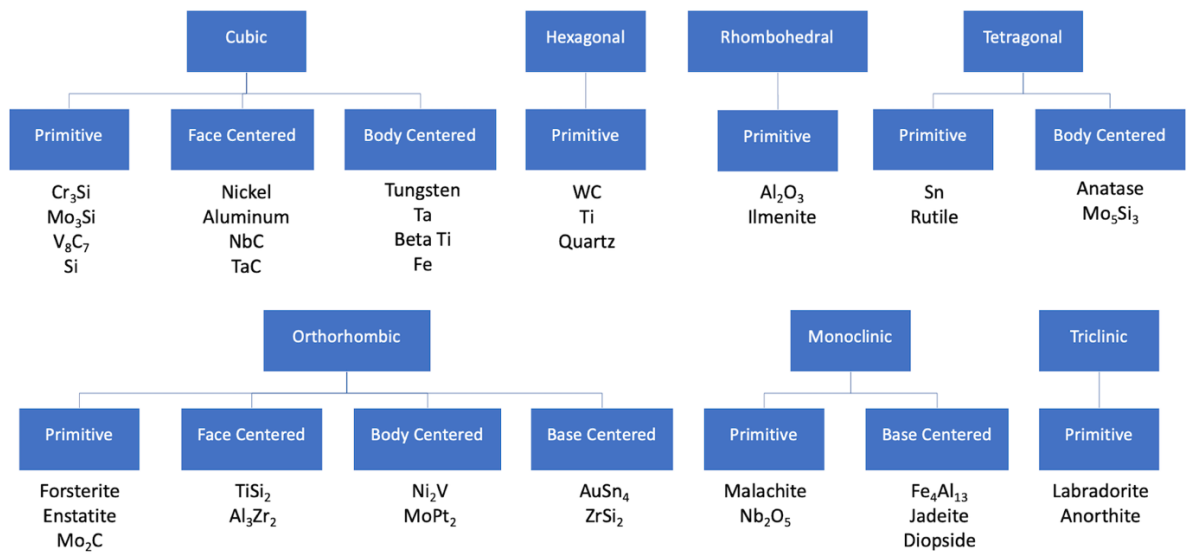

Extended Data Fig. 5 | Diagram of the materials and their Bravais lattice. For each of the fourteen Bravais lattices, diffraction patterns from at least two materials were collected. Diffraction patterns from supplementary materials in each Bravais lattice were utilized to test the performance of the model without it having any prior knowledge of the material.

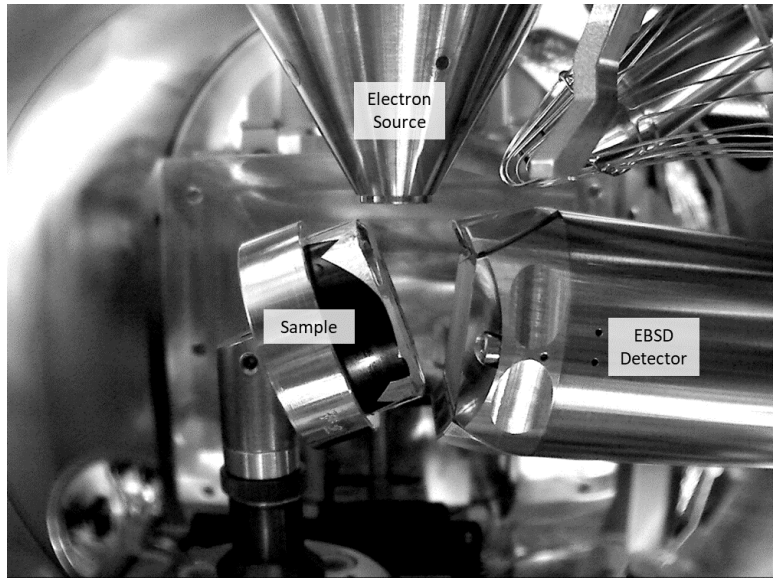

Extended Data Fig. 6 | Annotated image of the experimental setup within the SEM chamber. The sample is at 70 degrees from horizontal and is facing the EBSD detector. Some of the electrons which enter the sample may backscatter and exit the sample at the Bragg condition of the periodic atomic lattice planes. These electrons diffract to form Kikuchi bands corresponding to each of the lattice diffracting crystal planes. Most commercial systems then use look-up tables of user selected phases for orientation indexing.

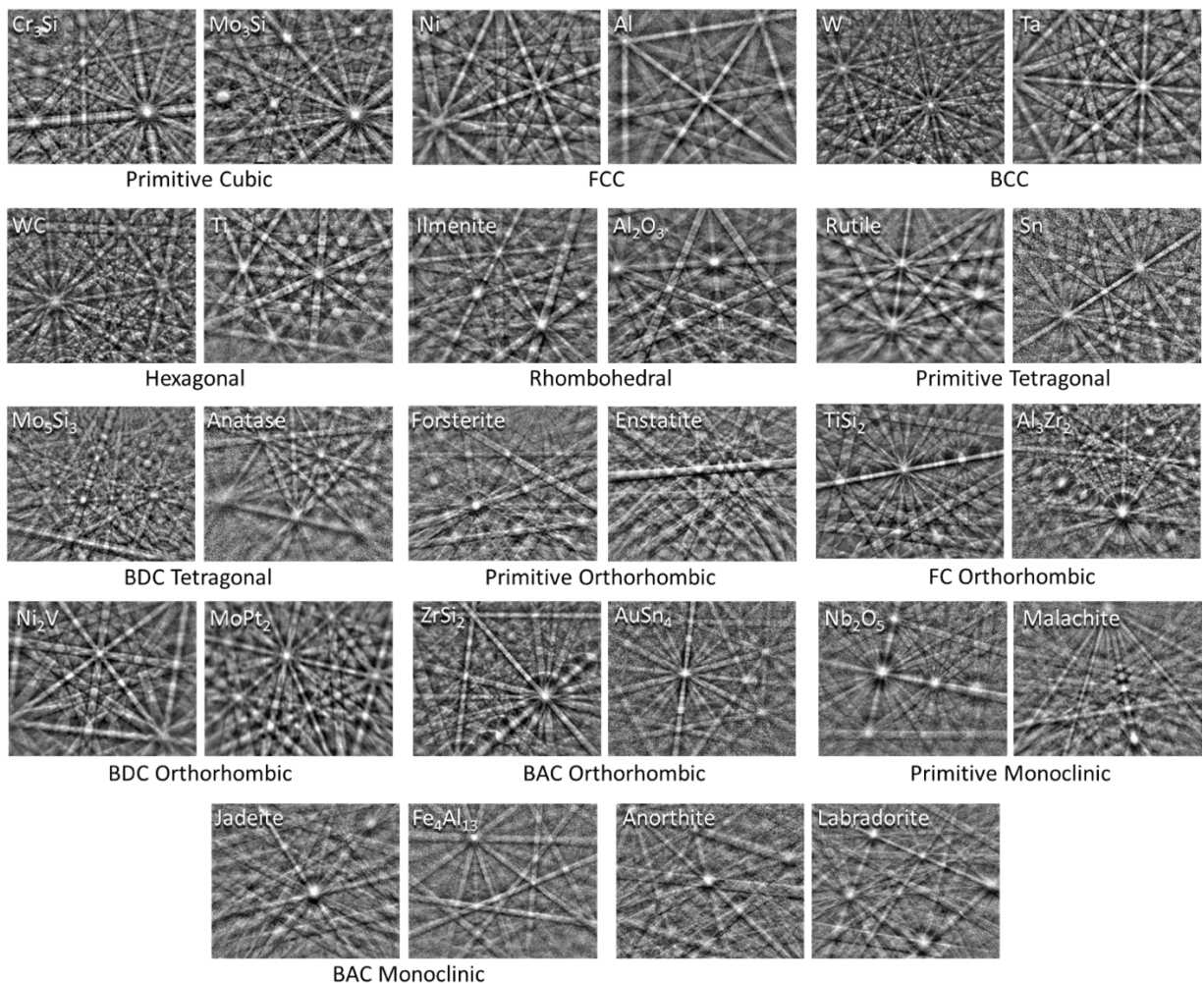

Extended Data Fig. 7 | An example diffraction pattern for each of the 28 materials used for training the machine learning model. This represents just one of the infinite number of orientations, and therefore diffraction patterns, the crystal can occupy in three-dimensional space.

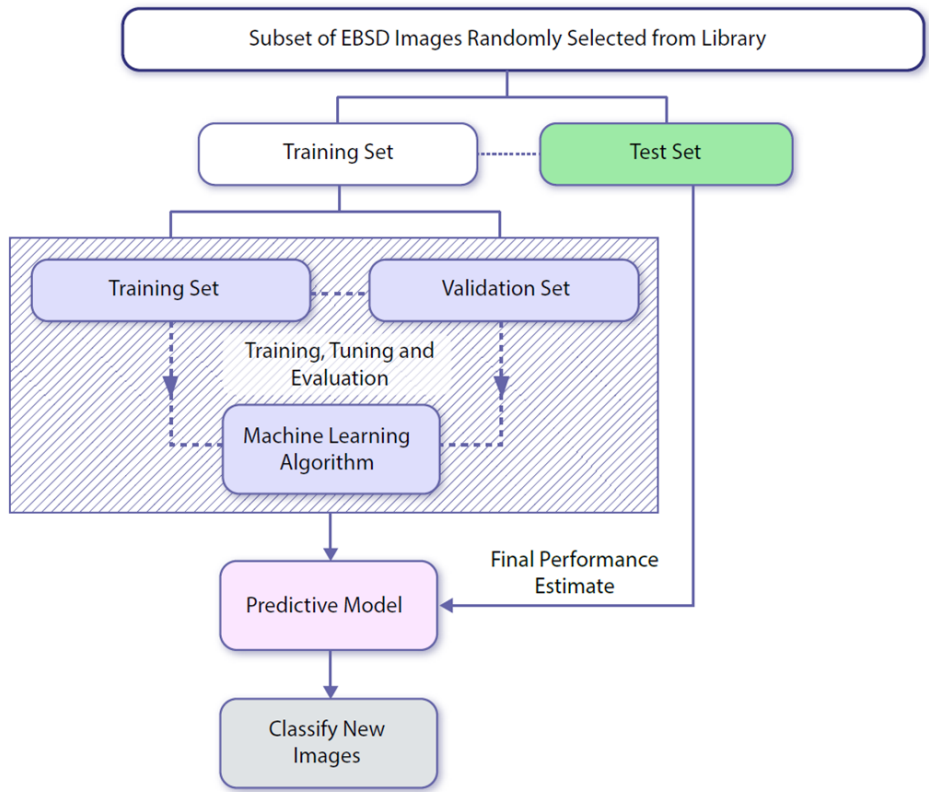

Extended Data Fig. 8 | Schematic describing the process of training a machine learning algorithm to recognize the symmetry features present in diffraction patterns. In the presented approach, a small subset of the EBSD patterns were randomly selected from two materials per class and then randomly subdivided into a training and testing set. The training set is then randomly divided again into a training set from which the machine learning model extracts features, and a validation set is used to test the usefulness of the features in correctly classifying the diffraction patterns. The feature bank is constantly tuned and evaluated such that the learned features produce the highest accuracy in classifying the validation set. When the machine learning algorithm determines it has found the best features for accurate classification, the learned filters are saved, and the predictive model can be applied to the classification of new diffraction patterns.

**7 Crystal Systems**

**14 Bravais Lattices**

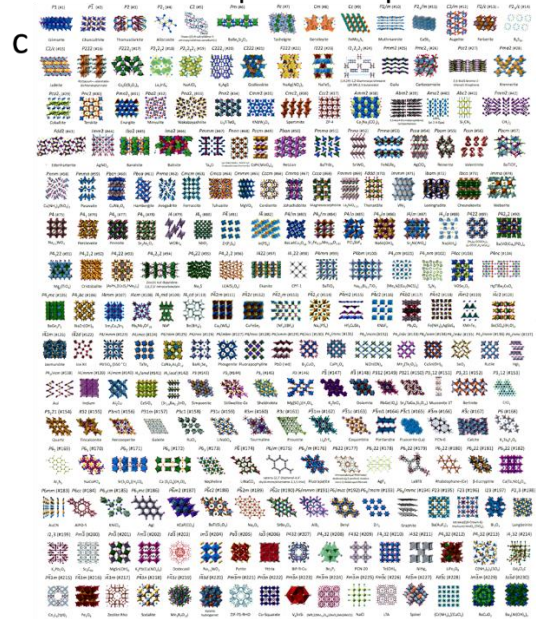

| Lattice System | Lengths | Angles | Primitive (P) | Base-centered (C) | Body-centered (I) | Face-centered (F) |
|---|---|---|---|---|---|---|
| Cubic | $a = b = c$ | $\alpha = \beta = \gamma = 90°$ | | | | |
| Hexagonal | $a = b \neq c$ | $\alpha = \beta = 90°$ $\gamma = 120°$ | | | | |
| Rhombohedral | $a = b = c$ | $\alpha = \beta = \gamma \neq 90°$ | | | | |
| Tetragonal | $a = b \neq c$ | $\alpha = \beta = \gamma = 90°$ | | | | |
| Orthorhombic | $a \neq b \neq c$ | $\alpha = \beta = \gamma = 90°$ | | | | |
| Monoclinic | $a \neq b \neq c$ | $\alpha = \beta = 90° \neq \gamma$ | | | | |
| Triclinic | $a \neq b \neq c$ | $\alpha \neq \beta \neq \gamma$ | | | | |

**32 Point Groups**

| Class | Group names | | | | | |
|---|---|---|---|---|---|---|
| Cubic | 23 | $m\bar{3}$ | | 432 | $\bar{4}3m$ | $m\bar{3}m$ |
| Hexagonal | 6 | $\bar{6}$ | 6/m | 622 | 6mm | $\bar{6}m2$ | 6/mmm |
| Trigonal | 3 | $\bar{3}$ | | 32 | 3m | $\bar{3}m$ |
| Tetragonal | 4 | $\bar{4}$ | 4/m | 422 | 4mm | $\bar{4}2m$ | 4/mmm |
| Orthorhombic | | | | 222 | | mm2 | mmm |
| Monoclinic | 2 | | 2/m | | m | |
| Triclinic | 1 | $\bar{1}$ | | | | |

**230 Space Groups**

Extended Data Fig. 9 | Crystal structure relationships from the seven crystal systems to the 230 space groups. **a,** Illustration of the fourteen Bravais lattices and their associated symmetries. **b,** A table of the 32 point groups and the crystal family they belong to. **c,** An example of each of the 230 space groups. Space groups represent the minute details of atomic arrangement within a point group. c) has been reproduced with permission from creator Frank Hoffmann under a creative commons (CC BY-NC-SA) license[57] and the crystal structures drawn with VESTA[58].